\documentclass[10pt, conference, letterpaper]{IEEEtran}
\usepackage[protrusion=true,expansion=true]{microtype}
\usepackage{microtype}
\usepackage[font=small]{caption}

\usepackage{cite}
\ifCLASSINFOpdf

\else

\fi
\usepackage{subfig}
\usepackage{bm}
\usepackage{amsmath}
\usepackage{epsf}
\usepackage{graphics}
\usepackage{ amssymb }
\usepackage[dvips]{graphicx}
\usepackage{epsfig}
\usepackage{cite}
\usepackage[linesnumbered,ruled,vlined]{algorithm2e}
\usepackage{colortbl}
\usepackage{color}
\usepackage{multicol}
\usepackage{bm}
\usepackage{amsmath}
\usepackage{epsf}
\usepackage{graphics}
\usepackage{ amssymb }
\usepackage[dvips]{graphicx}
\usepackage{epsfig}
\usepackage{cite}
\usepackage[linesnumbered,ruled,vlined]{algorithm2e}
\usepackage{graphicx}
\usepackage{epsfig}
\usepackage{latexsym}
\usepackage{amsfonts}
\usepackage{here}
\usepackage{rawfonts}
\usepackage{subfig}

\usepackage{bm}
\usepackage{amsmath}
\usepackage{epsf}
\usepackage{graphics}
\usepackage{ amssymb }
\usepackage[dvips]{graphicx}
\usepackage{epsfig}
\usepackage{cite}
\usepackage[linesnumbered,ruled,vlined]{algorithm2e}
\usepackage{colortbl}
\usepackage{color}
\usepackage{multicol}
\usepackage{bm}
\usepackage{amsmath}
\usepackage{epsf}
\usepackage{graphics}
\usepackage{ amssymb }
\usepackage[dvips]{graphicx}
\usepackage{epsfig}
\usepackage{cite}
\usepackage[linesnumbered,ruled,vlined]{algorithm2e}
\usepackage{graphicx}
\usepackage{epsfig}
\usepackage{latexsym}
\usepackage{amsfonts}
\usepackage{here}
\usepackage{rawfonts}

\usepackage[utf8]{inputenc}
\usepackage[english]{babel}
\usepackage{amsmath}
\usepackage{amsfonts}
\usepackage{amssymb}
\usepackage{color} 
\usepackage{bm}
\usepackage{listings}
\usepackage{caption}
\usepackage{amssymb}
\usepackage{amsthm}
\usepackage{graphicx}
\usepackage{epstopdf}
\usepackage{listings}
\usepackage{float}
\usepackage{amsmath}
\usepackage{amssymb}
\usepackage{amsfonts}
\usepackage{epstopdf}

\usepackage{multirow}
\usepackage{amscd}
\usepackage{mathrsfs}
\usepackage{graphicx}
\usepackage{color}
\usepackage{url}
\usepackage{bm}
\usepackage{setspace}

\usepackage{footnote}
\usepackage[utf8]{inputenc}
\usepackage[english]{babel}
\usepackage{amsmath}
\usepackage{amsfonts}
\usepackage{amssymb}
\usepackage{color} 
\usepackage{bm}
\usepackage{listings}
\usepackage{caption}
\usepackage{amssymb}
\usepackage{amsthm}
\usepackage{graphicx}
\usepackage{epstopdf}
\usepackage{listings}
\usepackage{float}
\usepackage{amsmath}
\usepackage{amssymb}
\usepackage{amsfonts}
\usepackage{epstopdf}

\usepackage{multirow}
\usepackage{amscd}
\usepackage{mathrsfs}
\usepackage{graphicx}
\usepackage{color}
\usepackage{url}
\usepackage{bm}
\usepackage{setspace}

\usepackage{footnote}
\usepackage[]{algorithm2e}
\renewcommand{\paragraph}{\@startsection{paragraph}{4}{\z@}%
	{-2.00ex\@plus -1ex \@minus -.2ex}%
	{0.5ex \@plus .2ex}%
	{\normalfont\normalsize}}
	
 \usepackage[protrusion=true,expansion=true]{microtype}

\IEEEoverridecommandlockouts
\usepackage{cite}
\usepackage{amsmath,amssymb,amsfonts}
\usepackage{algorithmic}
\usepackage{graphicx}
\usepackage{textcomp}
\usepackage{xcolor}

\usepackage[]{algorithm2e}
\renewcommand{\paragraph}{\@startsection{paragraph}{4}{\z@}%
	{-2.00ex\@plus -1ex \@minus -.2ex}%
	{0.5ex \@plus .2ex}%
	{\normalfont\normalsize}}
\RequirePackage{fix-cm} 

\makeatletter 
\newcommand\semiHuge{\@setfontsize\semiHuge{22.72}{27.38}}

\graphicspath{{img/}}
\begin{document}

\title{Energy Efficiency of RSMA and NOMA in  Cellular-Connected mmWave UAV Networks }

\author{Ali Rahmati$^*$, Yavuz Yap{\i}c{\i}$^*$, Nadisanka Rupasinghe$^*$, \.{I}smail G\"{u}ven\c{c}$^*$, Huaiyu Dai$^*$, and Arupjyoti Bhuyan$^\dagger$\\
$^*$Dept. of Electrical and Computer Engineering, North Carolina State University, Raleigh, NC\\
$^\dagger$Idaho National Laboratory, Idaho Falls, ID\\
Email:{\tt  \{arahmat,yyapici,rprupasi,iguvenc,hdai\}@ncsu.edu, arupjyoti.bhuyan@inl.gov}\thanks{This work is supported in part through the INL Laboratory Directed Research \& Development (LDRD) Program under DOE Idaho Operations Office Contract DE-AC07-05ID14517.}
}

\maketitle

\begin{abstract}
 Cellular-connected unmanned aerial vehicles (UAVs) are recently getting significant attention due to various practical use cases, e.g., surveillance, data gathering, purchase delivery, among other applications. Since UAVs are low power nodes, energy and spectral efficient communication is of paramount importance. To that end, multiple access (MA) schemes can play an important role in achieving high energy efficiency and spectral efficiency. In this work, we introduce rate-splitting MA (RSMA) and non-orthogonal MA (NOMA) schemes in a cellular-connected UAV network. In particular, we investigate the energy efficiency of the RSMA and NOMA schemes in a millimeter-wave (mmWave) downlink transmission scenario. Furthermore, we optimize precoding vectors of both the schemes by explicitly taking into account the 3GPP antenna propagation patterns. The numerical results for this realistic transmission scheme indicate that RSMA is superior to NOMA in terms of overall energy efficiency.    
\end{abstract}


\section{Introduction}
The use of unmanned aerial vehicles (UAVs) in wireless  networks is considered  a key component of   next generation communications systems~\cite{parvez2018survey, rahmati2019dynamic, mozaffari2019beyond}. Beside improving the network coverage through broadband mobile data delivery, UAVs have already become one of the major building blocks of various applications involving public safety, disaster relief, and surveillance, to name a few. In order to speed up deployment of UAVs in communications scenarios, one common approach is to employ UAVs as an integral part of the terrestrial wireless networks. By this way, infrastructure costs associated with serving flying UAVs are aimed to decrease as much as possible by making use of the existing network capabilities.   

In the \textit{cellular-connected UAV} concept, existing terrestrial wireless networks serve flying UAVs while their primary users are placed on the ground. The terrestrial BSs therefore have long-term relatively static features (e.g., down-tilted transmit antenna arrays) which are optimized considering ground users. The propagation pattern can, however, be \textit{shaped} taking into account not only the ground users but also the desired UAVs through advanced beamforming techniques. A more viable approach in simultaneously serving UAVs and ground users is to use state-of-the-art multiple access (MA) transmission schemes. By adequately allocating available  \textcolor{black}{resources} (e.g., power, time-frequency slots), the interference that arises when serving multiple users simultaneously  can be handled effectively through novel MA techniques. 

Non-orthogonal MA (NOMA) is identified as a promising MA scheme for 5G and beyond cellular communication technologies which can also be a smart solution for cellular-connected UAV scenarios~\cite{NadisankaTCoM, rupasinghe2018angle}. The NOMA transmission enables serving multiple UAVs at the same time, frequency and space resources enhancing the spectral efficiency. However, NOMA may not be the optimal MA scheme for multiple-input single-output (MISO) broadcast channels, except under overloaded conditions (i.e., user channels are highly correlated and not orthogonal). Moreover, the optimal user pairing and decoding order are two main drawbacks of the NOMA strategy in practical applications, which increase the system complexity. \textcolor{black}{Space division MA (SDMA) on the other hand is more suitable for underloaded conditions which however achieves degraded performance when network load is high.} \textcolor{black}{Recently, rate splitting MA (RSMA) is receiving significant attention as an effective MA scheme for next generation wireless communication systems.}  RSMA is preferable irrespective of the loading condition. In fact, RSMA enables soft bridging between two extremes, NOMA and SDMA~\cite{mao2018rate}.

In this work, we investigate the energy efficiency of various MA schemes in a downlink mmWave cellular-connected UAV network. In particular, we consider RSMA and NOMA as two MA schemes of interest, and optimize each scheme to maximize the energy efficiency.  Furthermore, we assume that the BS serving the desired UAVs actually belongs to a next generation terrestrial wireless network, and that the transmit antenna array is therefore composed of antenna elements having 3GPP propagation patterns. To the best of our knowledge, this is the first time that the energy efficiency of MA schemes are being studied in the context of cellular-connected mmWave UAV network with realistic propagation patterns. The numerical results verify the superiority of RSMA to NOMA in terms of energy efficiency in different scenarios.   



The rest of the paper is organized as follows.  In Section~\ref{sysmo}, the system model along with mmWave channel model and 3GPP antenna patterns are presented. The MA schemes are considered in Section~\ref{MA} together with exact energy efficiency optimization for RSMA and NOMA techniques involving 3GPP antenna patterns. Section~\ref{sim} presents numerical results on the performance of the considered schemes, and Section~\ref{con} finally concludes the paper.

\section{System Model}\label{sysmo}
In this section, we overview the downlink communication scenario involving UAVs as user terminals, mmWave channel model, and 3GPP antenna pattern considered in this work.

\subsection{Scenario Description}
\begin{figure}[!t]
	\includegraphics[width=0.48\textwidth]{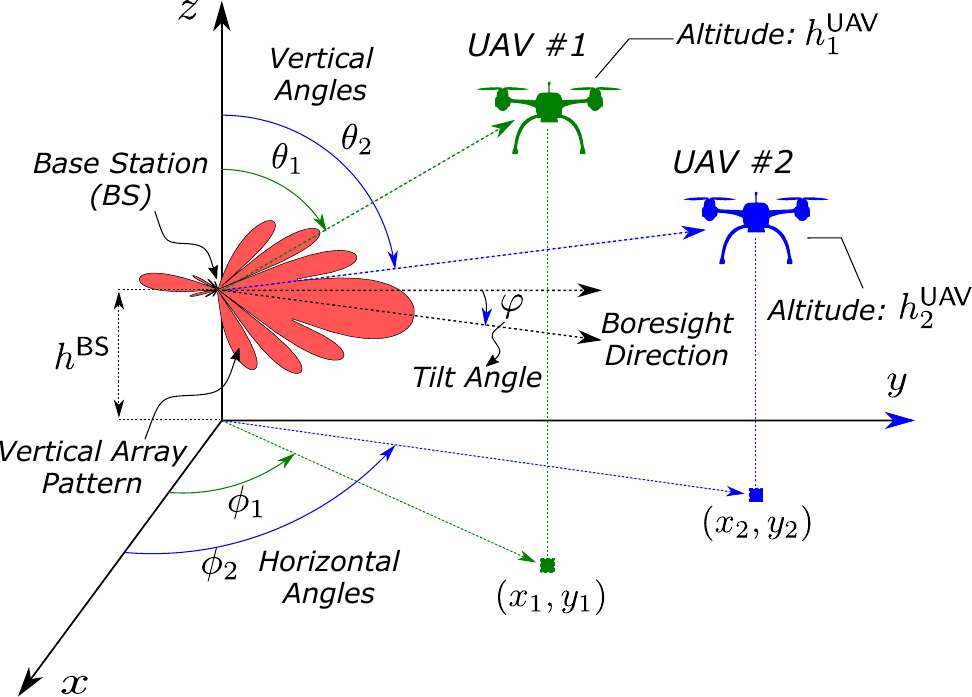}
	\centering
	\caption{System model for mmWave downlink serving cellular-connected UAVs simultaneously. The vertical propagation pattern of the $8$-element antenna array is also illustrated.}
	\label{fig:system}
\end{figure}

We consider a downlink transmission scenario in a single cell of a mmWave communications network. We assume that a conventional BS is responsible for serving desired ground users, and is therefore tilted downwards by an angle $\varphi$, as shown in Fig.~\ref{fig:system}. The transmit antenna structure at the BS is composed of a uniform linear array (ULA) with $M$ identical antenna elements placed vertically (i.e., along $z$-axis in Fig.~\ref{fig:system})~\cite{geraci2018understanding,3gpp2}. In our specific setting, we focus our attention to the downlink transmission where the BS is serving UAVs flying at various altitudes (i.e., \textit{cellular-connected UAVs}). Since the BS antenna array is configured to point downwards, the UAVs are assumed to be served using not only the main lobe but also the side lobes of the ULA propagation pattern depending on the altitude.

We assume that the terrestrial BS, which is off the ground vertically by $h^{\mathsf{BS}}$, is assigned a number of $K$ flying UAVs at altitudes $h_k^{\mathsf{UAV}}$ with the projection points $\left(x_k,y_k\right)$ on the $xy$-plane for $k\,{\in}\,\left\lbrace1,\dots,K \right\rbrace$. The BS aims to transmit dedicated messages to each UAV simultaneously (i.e., using the same time-frequency resources), where the message for the $k$th UAV is denoted by $W_k$. These messages are precoded based on various MA schemes, and form the
composite transmit signal $\textbf{{x}} \,{\in}\, \mathbb{C}^{M{\times}1}$. We assume that the total transmit power at the
BS is subject to a power constraint $P_t$ given as $\mathbb{E}\{||\textbf{x}||^2\} \,{\leq}\, P_t$. 

%

\subsection{mmWave Channel Model}
We assume a mmWave downlink transmission with $\mathbf{h}_k$ being the channel between the BS and the $k$th UAV, which is given as~\cite{Rappaport2017OveMil, lee2016randomly, ding2017random}
\begin{align}\label{eqn:channel}
\mathbf{h}_k=\sqrt{M} \sum_{p=1}^{N_p}\frac{\alpha_{k,p} \, \mathbf{a}(\theta_{k,p})}{\left[\textrm{PL}\left(\sqrt{d_k^2+\left(\Delta h_k^{\mathsf{UAV}}\right)^2}\right)\right]^{1/2}},
\end{align} 
where $N_P$ is the number of multipaths, $\alpha_{k,p}$ is the gain of the $p$th following standard complex Gaussian distribution, $\theta_{k,p}$ is the angle-of-departure
(AoD) of the $p$th path. Furthermore, $\Delta h_k^{\mathsf{UAV}} \,{=}\, h_k^{\mathsf{UAV}} \,{-}\, h^{\mathsf{BS}}$ describes the relative altitude of the $k$th UAV (i.e., vertical distance with respect to the BS) assuming $h_k^{\mathsf{UAV}} \,{>}\, h^{\mathsf{BS}}$, and $d_k \,{=}\, \sqrt{x_k^2 \,{+}\, y_k^2}$ stands for the ground distance of the $k$th UAV on the $xy$-plane (i.e., horizontal distance with respect to the BS). In addition, $\textrm{PL}$ represents the path loss of the downlink channel, which is described for the $k$th UAV using the line-of-sight (LoS) distance with respect to the BS. 

In \eqref{eqn:channel}, $\textbf{a}(\theta_{k,p})$ is the array steering vector, which is   
\begin{align}
\Big[ \textbf{a}(\theta_{k,p}) \Big]_i =  {\rm exp} \left\lbrace {-}j2\pi \frac{D}{\lambda} \left( i{-}1\right) \cos\left( \theta_{k,p} \right) \right\rbrace,
\end{align}
for $i \,{=}\, 1,\dots,M$ with $[\cdot]_i$ denoting the $i$th entry, where $D$ is the antenna element spacing along ULA, and $\lambda$ is the wavelength of the carrier frequency.  
We also assume that each UAVs has LoS path to the BS owing to sufficiently high flying altitudes, and the fact that probability of having scatters around any UAV is very small. Considering the well-known characteristic of the mmWave transmission where LoS path is significantly dominant as compared to the Non-LoS (NLoS) paths, we assume a single LoS path for the channel under
consideration, and \eqref{eqn:channel} accordingly becomes~\cite{NadisankaTCoM }
\begin{align}\label{eqn:channel_LoS}
\mathbf{h}_k=\sqrt{M}\frac{\alpha_{k} \, \mathbf{a}(\theta_{k})}{\left[\textrm{PL}\left(\sqrt{d_k^2+\left(\Delta h_k^{\mathsf{UAV}}\right)^2}\right)\right]^{1/2}},
\end{align}
where $\theta_k$ is the AoD of the LoS path. 

\subsection{Impact of 3GPP Antenna Pattern}

In the cellular-connected UAV scenario considered in our study, we assume that the transmit antenna array at the terrestrial BS is physically tilted downwards to serve the primary users of the existing network on the ground (i.e., see the angle $\varphi$ in Fig.~\ref{fig:system}). The desired UAVs are therefore served by not only the main lobe but also the side lobes of the antenna array propagation pattern \cite{mei2018uplink}. Hence, the specific pattern of the array antennas in the vertical domain plays a crucial role in the overall network performance. In our study, we adopt 3GPP realistic antenna pattern~\cite{rebato2018study, geraci2018understanding,3gpp2}, and present how to incorporate this model into our setting in the following.  

In order to weigh the transmission using the propagation pattern at the BS, we need to determine the vertical and horizontal angles $\theta$ and $\phi$, respectively. Considering the engagement geometry depicted in Fig.~\ref{fig:system}, the vertical angle for the $k$th UAV can be given as 
\begin{equation}\label{eqn:theta_k}
    \theta_k = \frac{\pi}{2} - \tan^{{-}1} \left(\frac{\Delta h_k^{\mathsf{UAV}}}{d_k}\right),
\end{equation}
and the horizontal angle is similarly given as
\begin{equation}\label{eqn:phi_k}
    \phi_k=\tan^{-1}\left(\frac{y_k}{x_k}\right).
\end{equation}

Considering 3GPP model~\cite{3gpp2, 3gpp}, radiation pattern of the single antenna element in the vertical domain is given as 
\begin{align}\label{eqn:pattern_vertical}
    A_{\mathsf{E,H}}(\theta,\varphi)= - \min \left\lbrace 12\left(\frac{\theta-90-\varphi}{\theta_{\textrm{3dB}}}\right)^2 \!\!, {SLA}_\mathsf{V} \right\rbrace ,
\end{align}
where $\theta_{\textrm{3dB}} \,{=}\, 65^\circ$ is the $3$-dB vertical beamwidth, ${SLA}_\mathsf{V} \,{=}\, 30$ dB is the side lobe level limit (i.e., side lobes away from the main lobe by greater than ${SLA}_\mathsf{V}$ is discarded), and $\varphi$ is the vertical tilt angle depicted in Fig.~\ref{fig:system}. Similarly, the
horizontal pattern of the single antenna element is given as
\begin{align}\label{eqn:pattern_horizontal}
    A_{\mathsf{E,H}}(\phi)=-\min \left\lbrace 12 \left(\frac{\phi}{\phi_{\textrm{3dB}}}\right)^2\!\!, A_\mathsf{m} \right\rbrace,
\end{align}
where $\phi_{\textrm{3dB}} \,{=}\, 65^\circ$ is the $3$-dB horizontal beamwidth, and
$A_\mathsf{m} \,{=}\, 30\,\text{dB}$ is the front-back ratio (i.e., the gain difference between the back and main lobes). Combining \eqref{eqn:pattern_vertical} and \eqref{eqn:pattern_horizontal}, the 3D antenna element propagation pattern for the angle pair $(\theta,\phi)$ is given as
\begin{align}\label{eqn:pattern_3D}
    A_\mathsf{E}(\theta,\phi) = G_{\max}-\min \left\lbrace-\big[ A_\mathsf{E,V}(\theta,\varphi)+ A_\mathsf{E,H}(\phi)\big],A_\mathsf{m}\right\rbrace
\end{align}
where $G_{\max}$ is the maximum directional gain of the
antenna element in the transmit antenna array. We note that \eqref{eqn:pattern_3D} stands for only the antenna element propagation pattern adopting 3GPP model, and therefore does not consider any contribution from array gain. {\color{black} The gain $G_k$ in \eqref{eqn:channel} can then be given as $G_k \,{=}\, A_\mathsf{E}(\theta_k,\phi_k)$} where the angle pair $\theta_k$ and $\phi_k$ for the $k$th UAV are given in \eqref{eqn:theta_k} and \eqref{eqn:phi_k}, respectively. It should be noted that $A_\mathsf{E}(\theta_k,\phi_k)$  is in dB but needs to be converted to linear scale.


\section{Multiple Access for Cellular-Connected UAVs} \label{MA}

In this section, we consider two MA schemes (i.e, RSMA and NOMA) to serve several cellular-connected UAVs simultaneously, and optimize the both schemes for energy efficiency considering the user rates and the total power required. In our derivations, we explicitly consider the realistic antenna pattern of 3GPP along with the specific engagement geometry.

\subsection{Rate Splitting Multiple Access (RSMA)}

The RSMA strategy is pioneered by \cite{rimoldi1996rate}, where it is shown that the achievable rate region obtained through successive interference cancellation (SIC) is just a fraction of its complete capacity region. In order to produce any rates in the capacity region as much as possible, the RSMA approach is proposed for MA scenarios. In this approach, instead of assigning message of each user to separate data streams, RSMA suggests to transmit a ``\textit{common}'' part of all user messages jointly, and the rest of each user message through different streams called ``\textit{private}''. At the receiver side, each user is expected to decode first the common message (in the presence of all private messages), cancel the effect of this common message from the received signal, and decode its own private message (in the presence of the other private messages).

We assume that the BS encodes part of each stream to be transmitted to a particular UAV as the \textit{common message}, denoted by $s_{\rm c}$. The rest of the stream for each UAV is then considered as \textit{private message}, and is encoded into stream $s_k$ for the $k$th UAV. In order to mitigate the interference as much as possible, we consider different beamforming vectors for each private message as well as the common message. Assuming that $\mathbf{w}_k$ and $\mathbf{w}_{\rm c}$ represent the beamforming vectors for the private message of the $k$th UAV and the common message, respectively, the overall beamforming matrix can be given as $\mathbf{W} \,{=}\, [\mathbf{w}_{\rm c} \, \mathbf{w}_1 \, \dots \, \mathbf{w}_K]$. In addition, the transmit signal at the BS after precoding is given as $\mathbf{x} \,{=}\, \mathbf{W} \mathbf{s}$ where $\mathbf{s} \,{=}\, [s_{\rm c} \, s_1 \, \dots \, s_K]$ is the overall message vector.

The received signal at the $k$th UAV is given by: 
\begin{align}
    y_k &= \sqrt{G_k}\mathbf{h}^H_k\mathbf{W}\mathbf{s} + n_k , \\
    &= \sqrt{G_k}\mathbf{h}^H_k\mathbf{w}_{\rm c} s_{\rm c} + \sqrt{G_k} \sum_{\ell=1}^{K}\mathbf{h}^H_k\mathbf{w}_\ell s_\ell + n_k,
\end{align}
where $G_k$ is the gain of the antenna element propagation pattern, $\mathbf{h}^{\rm H}_k\,{\in}\, \mathbb{C}^{M{\times}1}$ is the channel from the BS to the $k$th UAV, which is assumed to be known perfectly at the BS in our study, and $n_k$ is the circularly symmetric complex Gaussian white noise with zero-mean
and variance $N_0$, i.e., $n_k\,{ \sim}\, \mathcal{CN}(0,N_0)$.
Assuming unit-energy message vector $\textbf{s}$, the power budget at the BS requires ${\rm tr}(\mathbf{W}\mathbf{W}^H) \,{\leq}\, P_\textrm{t}$, which needs to be handled properly in the optimization problem.

In order to decode the entire message for each UAV, the common message has to be decoded first in the presence of the private messages of \textit{all} the UAVs, and then cancel its effect from the received symbol, which is known as successive interference cancellation (SIC). Assuming that common message is successfully decoded in all the UAVs, each UAV then decodes its own private message treating the other private messages as noise. The SINR of the $k$th UAV while decoding the common message $s_{\rm c}$ is given as 
\begin{equation}
    \mathsf{SINR}^{\rm c}_{k} = 
\frac{G_k \left|\mathbf{h}^H_k \mathbf{w}_{\rm c} \right|^2}
{G_k \sum_{\ell=1}^K \left| \mathbf{h}^H_k\mathbf{w}_\ell \right|^2 + N_{0}},
\end{equation}
and the respective rate (normalized by transmission bandwidth) is expressed as $\mathsf{R}^{\rm c}_{k} =
\log_2 \left(1 + 
\mathsf{SINR}^{\rm c}_{k} \right)$.
We note that the overall success of the decoding process involves private message as well as common message being decoded without any error \textit{at each} UAV. The necessary condition for successful decoding of the common message is that the respective common rate $r^{\rm c}_k$ for the $k$th UAV should be less than $\mathsf{R}^{\rm c} \,{=}\,
\min \{\mathsf{R}^{\rm c}_{1}, \dots, \mathsf{R}^{\rm c}_{K} \}$. 

After successfully decoding the common message, and cancelling its contribution from the received signal, each UAV can decode its own private message in the presence of the other private messages. The respective SINR for the $k$th UAV is
\begin{align}
    \mathsf{SINR}^{\rm p}_{k} = 
\frac{G_k \left|\mathbf{h}^H_k \mathbf{w}_k \right|^2}
{G_k \sum_{\ell\neq k} \left| \mathbf{h}^H_k\mathbf{w}_\ell \right|^2 + N_{0}},
\end{align}
and the normalized rate is $\mathsf{R}^{\rm p}_{k} =
\log_2 \left(1 + 
\mathsf{SINR}^{\rm p}_{k} \right)$.

In this study, we optimize the precoding matrix $\textbf{W}$ and the common rate vector $\mathbf{r}^{\rm c} \,{=}\, [r^{\rm c}_{\rm 1} \, \dots \, r^{\rm c}_K]$ to maximize the energy efficiency, which is given for the $k$th UAV as
\begin{align}\label{eqn:ee_rsma_k}
    \mathsf{EE}^{\mathsf{RSMA}}_k =  \frac{\beta_k \left( r^{\rm c}_k + \mathsf{R}^{\rm p}_{k} \right) } { {\color{black} {\rm tr}( \mathbf{W} \mathbf{W}^H )} + P_\mathsf{BS}},
\end{align}
where $P_\mathsf{BS}$ is the fixed power consumption to operate the BS (even without RF transmission), and $\beta_k$ is the communication-dependent weight factor which might be used to prioritize UAVs based on their quality-of-service (QoS) requirements.
The energy efficiency for the overall network is
\begin{align}\label{eqn:ee_rsma}
    \mathsf{EE}^{\mathsf{RSMA}} =  \sum_{k=1}^K \mathsf{EE}^{\mathsf{RSMA}}_k,
\end{align}
and the respective optimization problem is given as 
\begin{IEEEeqnarray}{rl}\label{eqn:opt_ee_rsma}
\max_{\mathbf{r}^{\rm c}, \textbf{W}}
&\qquad \mathsf{EE}^{\mathsf{RSMA}}  \\
\text{s.t.}
&\qquad {\rm tr} \left(\mathbf{W}\mathbf{W}^H \right) \leq P_\textrm{t} , \IEEEyessubnumber\\
&\qquad \sum\nolimits_{\ell=1}^{K} r^{\rm c}_\ell \leq \mathsf{R}^{\rm c}_{k} \;\; \forall k \in \{1,\dots,K\},  \IEEEyessubnumber\\
&\qquad \mathbf{r}^{\rm c} \geq \mathbf{0}, \IEEEyessubnumber
\end{IEEEeqnarray}
where the optimization runs over both the common message rate vector $\textbf{r}^{\rm c}$ as well as the beamforming matrix $\textbf{W}$.


\subsection{Non-Orthogonal Multiple Access (NOMA)}


The NOMA strategy relies on the fact that user messages are \textit{distinct} enough in the power domain, and each receiver can decode its desired message using the SIC approach. To this end, the NOMA transmitter allocates adequate power to each UAV's message, which requires the knowledge of the channel condition for each UAV. Without any loss of generality, the UAVs with indices from $\mathcal{N} \,{=}\, \{1,\dots,K\}$ are ordered from the one allocated less power to the one allocated more power, and that BS has this information perfectly. Although the power allocation is jointly optimized in the sequel along with the beamforming matrix $\textbf{W}$, it is worth noting that the NOMA strategy proceeds with allocating less power to the UAVs having better channel conditions, and vice versa.  

Assuming that $\mathbf{w}_k$ is the beamforming vector for the $k$th UAV, the transmit signal at the BS after precoding is given as $\mathbf{x} \,{=}\, \mathbf{W} \mathbf{s}$ where $\mathbf{W} \,{=}\, [\mathbf{w}_1 \, \dots \, \mathbf{w}_K]$. The received signal at the $k$th UAV  now becomes 
\begin{align}\label{eqn:observations_noma}
    y_k &= \sqrt{G_k}\mathbf{h}^H_k\mathbf{W}\mathbf{s} + n_k = \sqrt{G_k} \sum_{\ell=1}^{K}\mathbf{h}^H_k\mathbf{w}_\ell s_\ell + n_k.
\end{align}
In the NOMA strategy, $k$th UAV is expected to first decode messages of the other UAVs allocated more power (i.e., with the index $m\,{>}\,k$). This decoding is performed by treating the UAVs allocated less power (i.e., with the index $m\,{<}\,k$) as noise. Each decoded message is then subtracted from the received signal in \eqref{eqn:observations_noma} through SIC, and $k$th UAV finally decodes its own message $s_k$. In this approach, the the SINR associated with decoding the message of the $m$th UAV (at the $k$th UAV with $m\,{\geq}\,k$) is given as
\begin{align} \label{eq:sinr_noma}
\mathsf{SINR}_{m{\rightarrow}k} = \frac{G_k|\textbf{h}_{k}^{\rm H}\textbf{w}_m|^2 }{G_k \sum \limits_{l < m,\, l \in \mathcal{N}}|\textbf{h}_{k}^{\rm H}\textbf{w}_\ell|^2 + N_0},
\end{align} 
which implicitly assumes that messages of all the UAVs with the indices greater than $m$ have been decoded successfully and canceled beforehand. Actually, \eqref{eq:sinr_noma} represents the SINR of the $k$th UAV while decoding its own message (i.e., for $m\,{=}\,k$), as well. Note that the summation in the denominator of \eqref{eq:sinr_noma} disappears when the $k$th UAV with $k\,{=}\,1$ is decoding its own message (i.e., no possible $l$ index satisfying $l \,{<}\, k$ for this specific case), and we have $\frac{G_k}{N_0}|\textbf{h}_{k}^{\rm H}\textbf{w}_k|^2$. 

Finally, the normalized rate associated with the $k$th UAV is
\begin{align}
\mathsf{R}_{k} = \log_2 \left( 1 +
\min\left\lbrace 
\mathsf{SINR}_{k{\rightarrow}1},\dots,\mathsf{SINR}_{k{\rightarrow}k}
\right\rbrace \right).
\end{align}
The energy efficiency for the $k$th UAV is \begin{align}\label{eqn:ee_noma_k}
    \mathsf{EE}^{\mathsf{NOMA}}_k = \frac{\beta_k \mathsf{R}_{k}} { {\color{black} {\rm tr}( \mathbf{W} \mathbf{W}^H )} + P_\mathsf{BS}},
\end{align}
and for the overall network is
\begin{align}\label{eqn:ee_noma}
    \mathsf{EE}^{\mathsf{NOMA}} = \sum_{k=1}^K \mathsf{EE}^{\mathsf{NOMA}}_k.
\end{align}
The respective optimization problem therefore becomes
\begin{IEEEeqnarray}{rl}\label{eqn:opt_ee_noma}
\max_{\mathbf{r}^{\rm c}, \textbf{W}}
&\qquad \mathsf{EE}^{\mathsf{NOMA}}  \\
\text{s.t.}
&\qquad {\rm tr} \left(\mathbf{W}\mathbf{W}^H \right) \leq P_\textrm{t} . \IEEEyessubnumber
\end{IEEEeqnarray}



\subsection{Successive Convex Approximation (SCA)}

The energy efficiency optimization problems for RSMA and NOMA given by \eqref{eqn:ee_rsma} and \eqref{eqn:ee_noma}, respectively, are highly non-convex. Thus, we need to come up with an approximate algorithm to solve these optimization problems efficiently. As in \cite{mao2018energy}, we adopt the SCA algorithm~\cite{7946258} as a good match for this problem. The algorithm proposes to approximate the non-convex  objective function (or the functions in the constraints) using the first-order Taylor expansion around a given initial point. By this way, the problem is converted to a convex optimization problem, and can therefore be solved efficiently using standard optimization toolboxes (e.g., CVX~\cite{grant2008cvx}, YALMIP~\cite{lofberg2004yalmip}). The algorithm proceeds through iterations such that the optimal values of the variables in the first iteration are employed as the initial values for the next iteration. This procedure is repeated until the convergence is achieved.

\section{Simulation Results}\label{sim}
In this section, we present numerical results for the performance of RSMA and NOMA considering a downlink mmWave network with $2$ cellular-connected UAVs. We adopt the path loss model in \cite{lee2016randomly, ding2017random} with $\textrm{PL}(x_k)=1+x_k^\gamma$, where $\gamma$ is the path loss exponent, and $x_k$ is the LoS distance between the BS and the $k$th UAV. We assume that the antenna array of the BS has $M\,{=}\,8$ elements, which are placed vertically. We assume that the UAVs are fixed in $xy$-plane during the simulation. The specific simulation parameters are listed in Table~\ref{tab:parameters}.  

\begin {table}[t!]
\vspace{8mm}
\caption {Simulation Parameters} \label{tab:parameters} 
\renewcommand{\arraystretch}{1.2}
\centering
\begin{tabular}{ lc }
\hline
Parameter & Value \\
\hline
\hline
Path loss exponent $(\gamma)$ & $2$  \\
BS height $(h^{\mathsf{BS}})$ & {\color{black}$10\,\text{m}$ } \\
UAV altitude $(h^{\mathsf{UAV}}_1, h^{\mathsf{UAV}}_2)$ & $10-60\,\text{m}$\\
BS array size $(M)$ & $8$ \\ 
Noise variance $(N_0)$ & $1$ \\ 
Total transmit power $(P_t)$ & $40\,\text{dBm}$ \\ 
Weight of $1$st UAV $(\beta_1)$ & 1 \\ 
Weight of $2$nd UAV $(\beta_2)$ & $10^\eta$, $\eta \in [-2,2]$ \\
Circuit power consumption $(P_{\textrm{c}})$ & $40\,\text{dBm}$ \\ 
Tilt angle $(\varphi)$ & $12^\circ$ \\ 
Maximum directional gain  $(G_{\max})$ & $ 8\,\text{dBi}$ \\ 
 \hline
\end{tabular}
\end {table}
\begin{figure}[!t]
	\centering
	\includegraphics[width=0.5\textwidth]{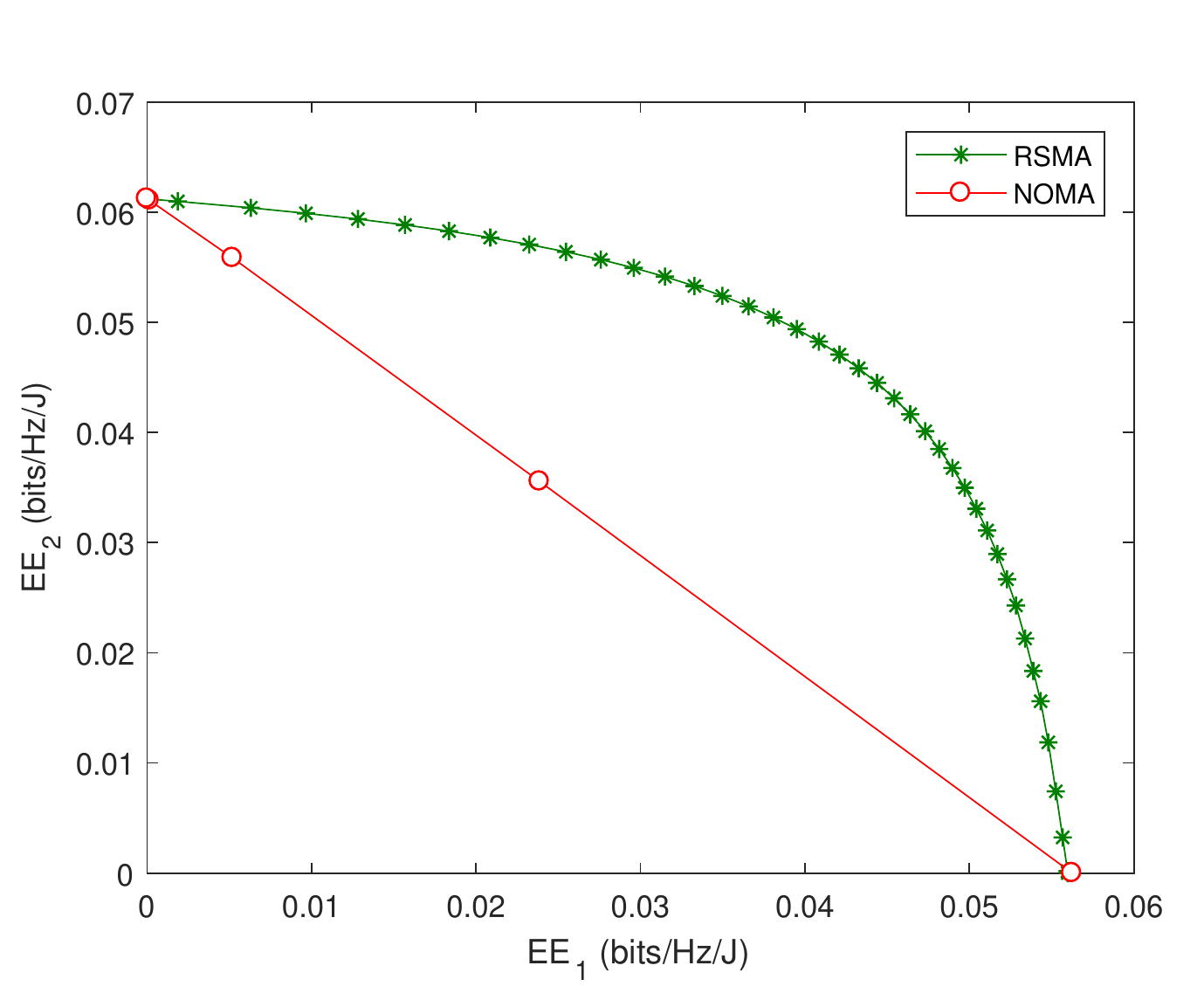}\vspace{-0.1in}
	\caption{Energy efficiency region for RSMA and NOMA.}
	\label{f2}
\end{figure}

\begin{figure}[!t]
	\centering
	\includegraphics[width=0.5\textwidth]{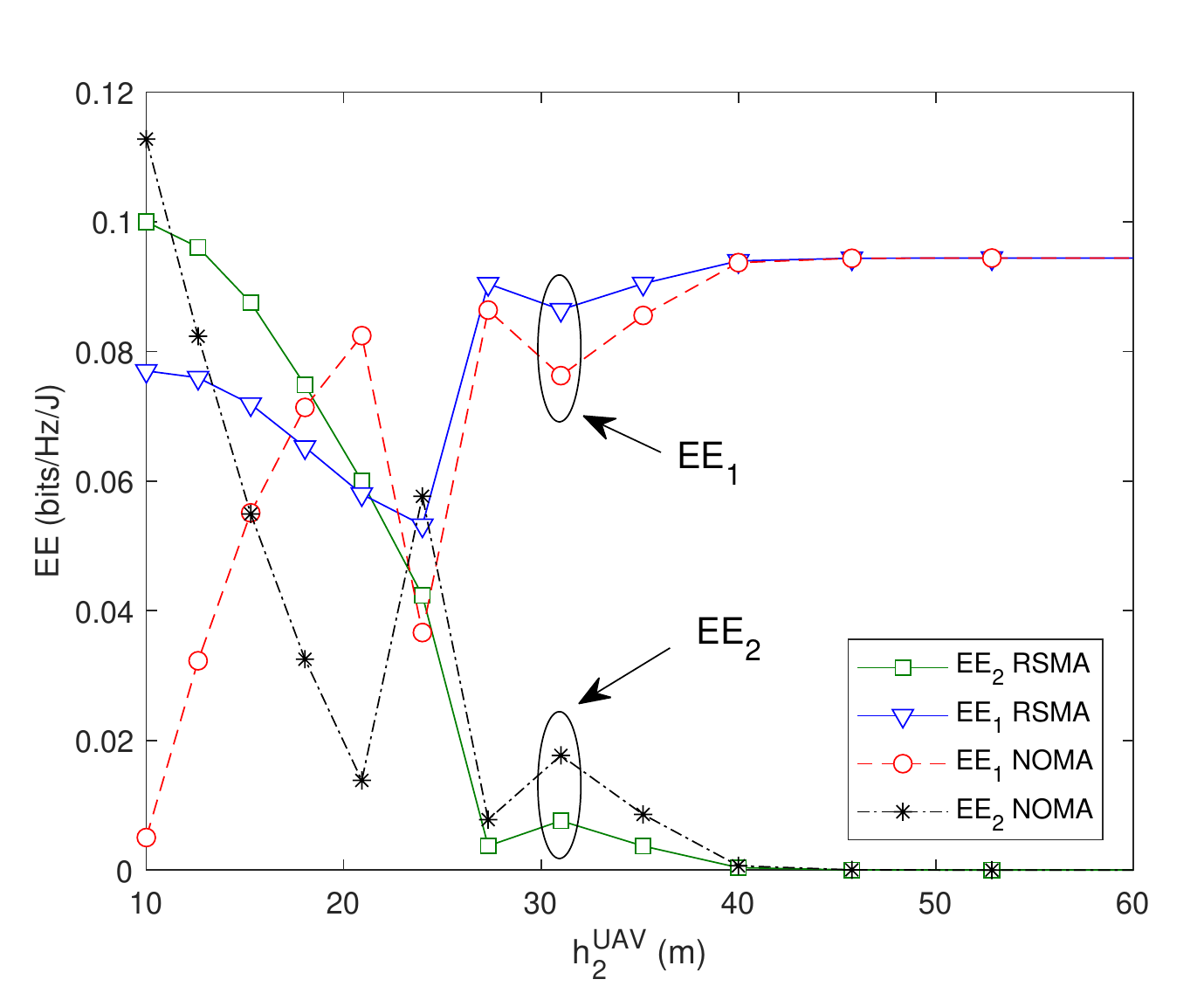}\vspace{-0.1in}
	\caption{Energy efficiency of each UAV versus the altitude of the $2$nd UAV.}
	\label{f3}
\end{figure}

\begin{figure}[!t]
	\centering
	\includegraphics[width=0.5\textwidth]{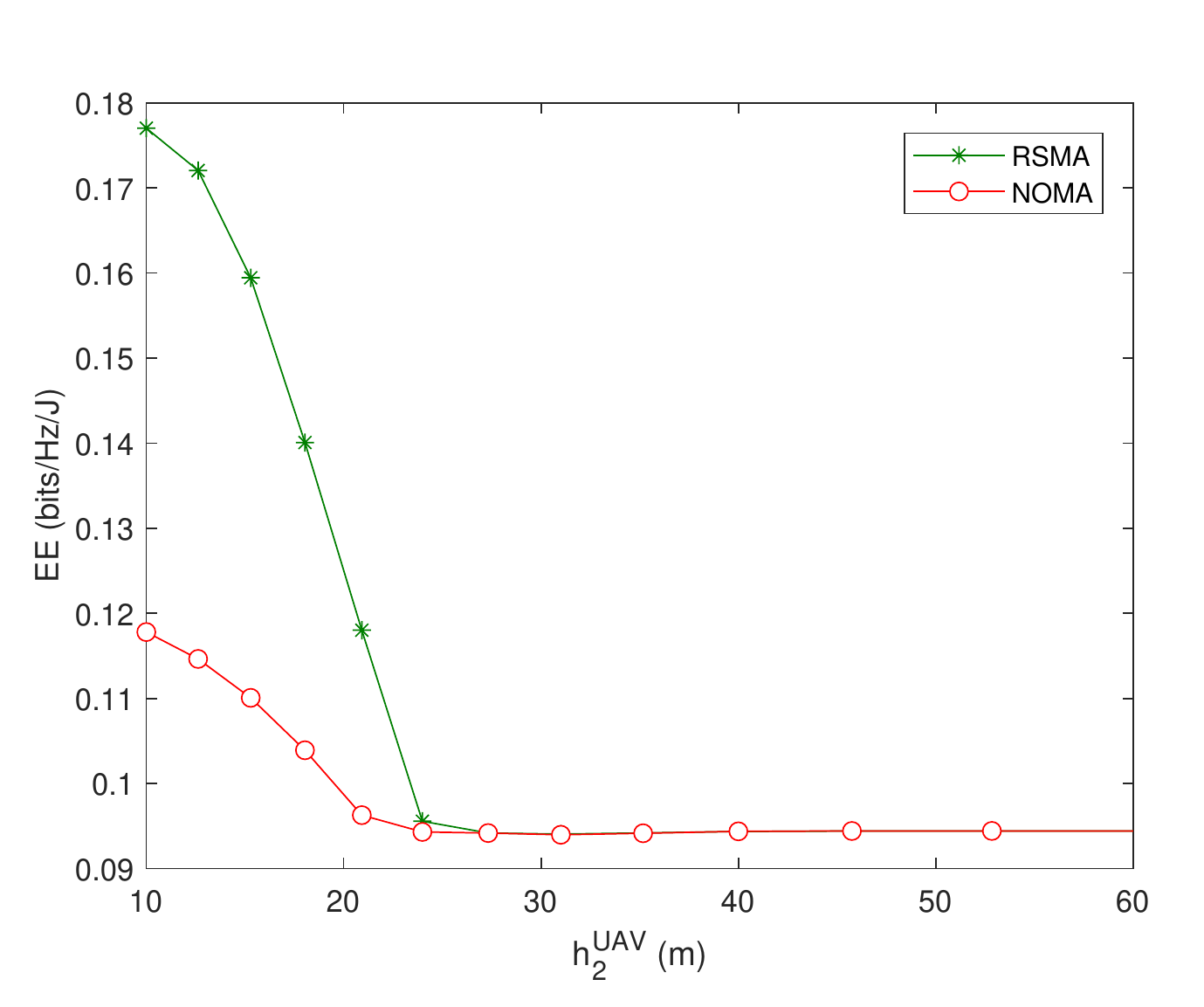}\vspace{-0.1in}
	\caption{Sum energy efficiency versus the altitude of the $2$nd UAV.}
	\label{f4}
\end{figure}
In Fig.~\ref{f2}, we depict the energy efficiency region of the $2$ cellular-connected UAVs, where the weight factor of the $1$st UAV is fixed as $\beta_1 \,{=}\, 1$, and that of the $2$nd UAV is varying as $\beta_2 \,{=}\, 10^\eta$ with $\eta \,{\in}\, [-2,2]$. We furthermore assume a geometry for the model of Fig.~\ref{fig:system} for which the locations of the UAVs on the ground are given by the horizontal distances $d_1 \,{=}\, 25\,\text{m}$ and $d_2 \,{=}\, 30\,\text{m}$, and angles $\phi_1 \,{=}\, \pi/10$, and $\phi_2 \,{=}\, 2\pi/5$. In addition, altitudes of the UAVs are given as {\color{black}$h_1^{\mathsf{UAV}} \,{=}\, 24.4\,\text{m}$} and {\color{black}$h_2^{\mathsf{UAV}} \,{=}\, 15.3\,\text{m}$}, which correspond to the vertical angles $\theta_1 \,{=}\, \pi/3$ and $\theta_2 \,{=}\, 4\pi/9$ by \eqref{eqn:theta_k}. We observe in Fig.~\ref{f2} that the energy efficiency of the UAVs with RSMA is superior to that for NOMA for any choice of weight factor $\beta_2$.  

In Fig.~\ref{f3}, we demonstrate the energy efficiency of each UAV along with varying altitude of the $2$nd UAV such that {\color{black}$h_2^{\mathsf{UAV}} \,{\in}\, [0,60]\,\text{m}$}. To this end, we keep the simulation parameters of Fig.~\ref{f2} the same for this experiment except $h_2^{\mathsf{UAV}}$, which requires the vertical angle to vary as {\color{black}$\theta_2 \,{\in}\, [\pi/18,\pi/2]$} by \eqref{eqn:theta_k}. The sum energy efficiency given by \eqref{eqn:ee_rsma} and \eqref{eqn:ee_noma} are also depicted in Fig.~\ref{f4}. We observe that the energy efficiency of the $2$nd UAV exhibits a decaying trend for both RSMA and NOMA, since it is not reasonable to allocate more power to the $2$nd UAV as its altitude gets larger, and, hence, the respective channel degrades along with increasing path loss. As a result, the $1$st UAV is allocated more power, and its energy efficiency improves as $h_2^{\mathsf{UAV}}$ increases. We also observe that although individual energy efficiency of RSMA and NOMA can be superior to one another at certain altitudes, and the sum energy efficiency of RSMA is significantly superior to that of NOMA.

\begin{figure}[!t]
	\centering
	\includegraphics[width=0.5\textwidth]{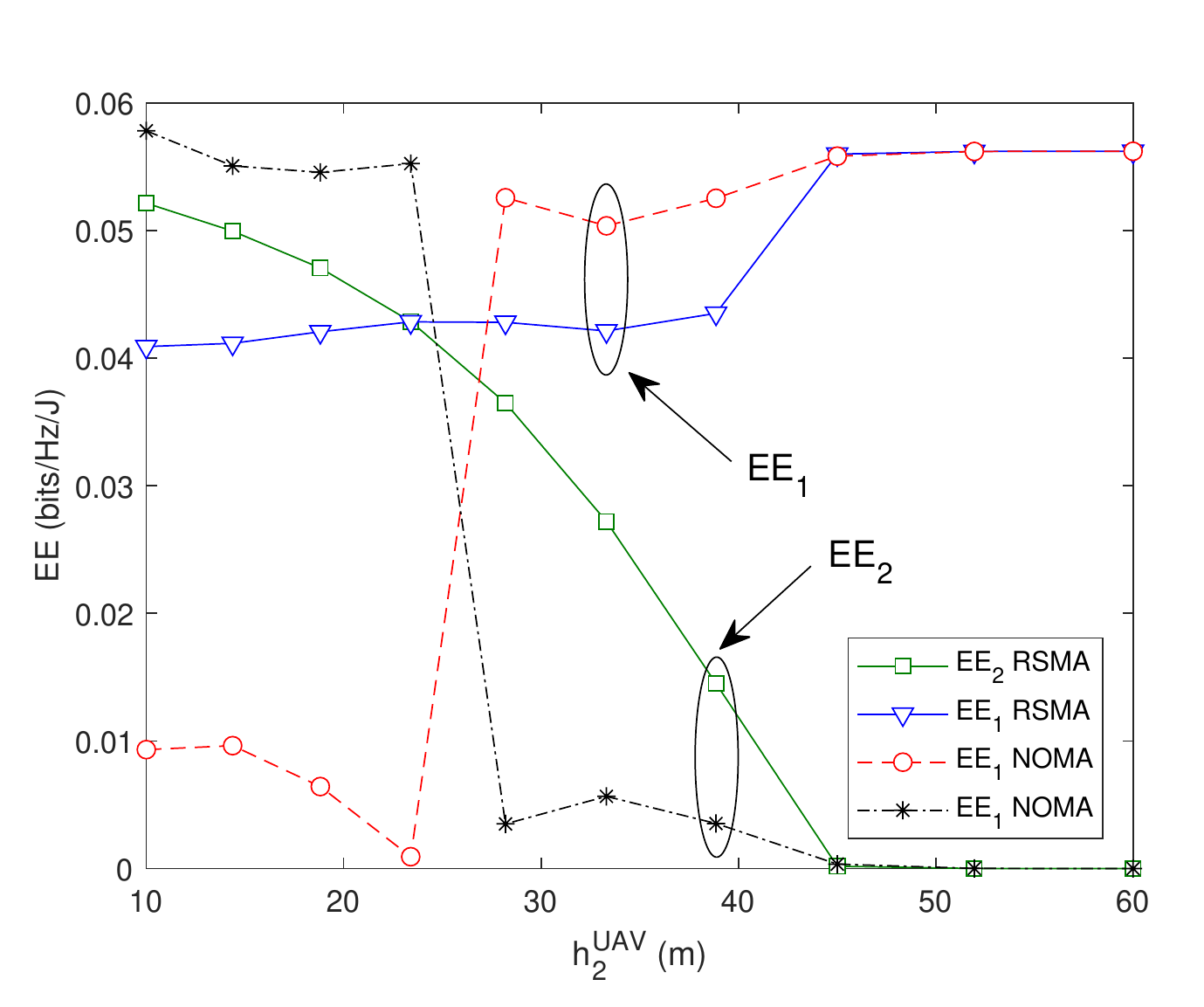}\vspace{-0.1in}
	\caption{Energy efficiency of each UAV versus the altitude of the $2$nd UAV.}
	\label{f6}\vspace{-0.2in}
\end{figure}

\begin{figure}[!t]
	\centering
	\includegraphics[width=0.5\textwidth]{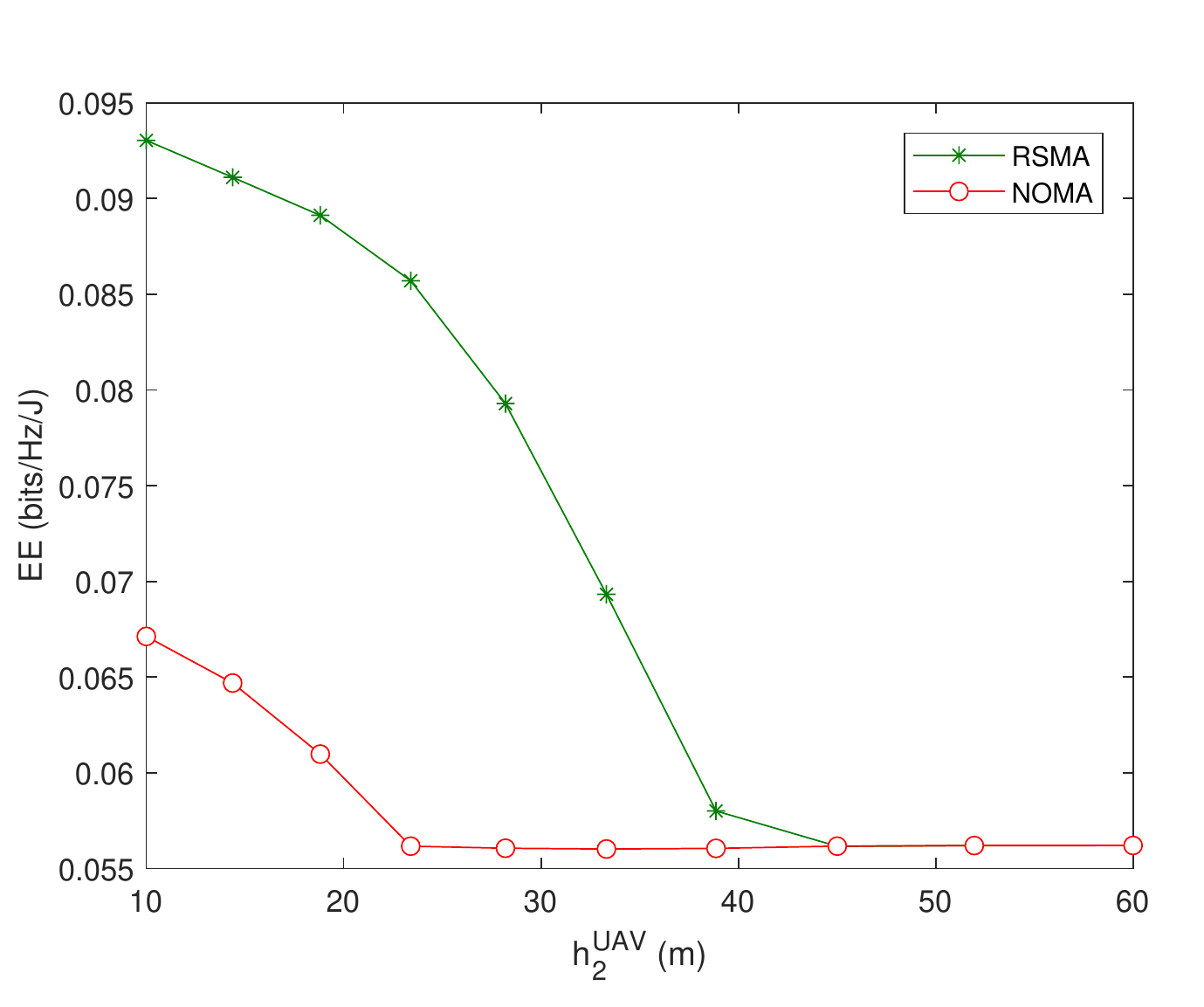}\vspace{-0.1in}
	\caption{Sum energy efficiency versus the altitude of the $2$nd UAV.}
	\label{f7}
\end{figure}


We now change the setting such that the locations of UAVs projected onto the ground are given by the horizontal distances $d_1 \,{=}\, 20\,\text{m}$ and $d_2 \,{=}\, 50\,\text{m}$, and angles $\phi_1 \,{=}\, \pi/10$, and $\phi_2 \,{=}\, 2\pi/5$. In addition, altitude of the $1$st UAVs is {\color{black}$h_1^{\mathsf{UAV}} \,{=}\, 14.4\,\text{m}$} with a vertical angle of $\theta_1 \,{=}\, 4\pi/9$, and altitude of $2$nd UAV is varying such that $h_2^{\mathsf{UAV}} \,{\in}\, [0,60]\,\text{m}$, as before. 
The respective individual and sum energy efficiency results are depicted in Fig.~\ref{f6} and Fig.~\ref{f7}, respectively. We observe that the performance gap of RSMA and NOMA appears to be more significant in favor of RSMA in this particular setting (as compared to that of Fig.~\ref{f4}). 

\section{Conclusion}\label{con}
We consider a downlink mmWave transmission in a cellular-connected UAV network. The desired UAVs are assumed to be served by a terrestrial BS, where the transmit array employs the 3GPP antenna propagation patterns. We also assume that the BS serves the desired UAVs simultaneously using either RSMA or NOMA as the MA scheme. We show that the energy efficiency of either RSMA or NOMA varies non-monotonically with the operation altitude, and   RSMA is superior to NOMA in terms of total energy efficiency 

\bibliographystyle{IEEEtran}
\bibliography{IEEEabrv,references}

\end{document}